\documentclass[pre, twocolumn, amsmath,amssymb]{revtex4-1}
\usepackage{graphicx}
\usepackage{amsfonts}
\usepackage{bm}
\usepackage{dcolumn}
\usepackage{color}
\begin{document}

\title{Minimal two-sphere model of the generation of fluid flow at low
   Reynolds numbers}

\author{M. Leoni$^{1,2,3}$}
\author{B. Bassetti$^1$}
\author{J. Kotar$^2$}
\author{P. Cicuta$^2$}
\author{M. Cosentino Lagomarsino$^1$}

\affiliation{$^1$ Dip. Fisica, Universit\`a di Milano, Via Celoria,
16, 20133 Milano, Italy, and INFN, Milano, Italy}
\affiliation{$^2$ Cavendish Laboratory and Nanoscience Centre, University of Cambridge, Cambridge CB3 0HE, U. K.}
\affiliation{$^3$ Department of Mathematics, University of Bristol, Clifton, Bristol BS8 1TW, U.K.}

\begin{abstract}
  Locomotion and generation of flow at low Reynolds number are subject to severe limitations due to the irrelevance of inertia: the  ``scallop theorem'' requires that the system have at least two degrees of freedom, which move in \emph{non-reciprocal} fashion, i.e. breaking time-reversal symmetry. We show here that a minimal model consisting of just two spheres driven by harmonic potentials is capable of generating flow. In this pump system the two degrees of freedom are the mean and relative positions of the two spheres. We have performed and compared analytical predictions, numerical simulation and experiments, showing that a \emph{a time-reversible} drive is sufficient to induce flow.\end{abstract}
\maketitle

Microscopic systems capable of generating flow are very common in Nature, and may prove inspirational for bio-mimetic micro- and nano-pumps and swimmers~\cite{Lau09}.  Assemblies of motile cilia are found in various eukaryotic living systems. In humans, for example, they cover the epithelial tissue of important organs, including the lungs, ventricles in the brain, and the oviduct in the female reproductive apparatus~\cite{Sher}. They transport fluid along their surface in a given direction by controlling the effective drag coefficient to change between a ``power''  and a ``recovery'' stroke~\cite{Bray}.  This is one simple way to satisfy the ``scallop theorem''~\cite{Pur77} that sets very strict physical requirements for swimming and pumping at small velocity to viscosity ratio at the microscale, where the Reynolds number is to good approximation zero. A necessary condition, in order to pump, is that the sequence of the system's configurations has to break time-reversal symmetry~\cite{Pur77}. The scallop theorem applies to pumps as well as swimmers~\cite{RA07}, so no net flow will occur unless the generating motion is \emph{non-reciprocal}. This implies a minimum of two degrees of freedom, with which time-reversal symmetry can be broken by an appropriate sequence of moves~\cite{NG04,LE+09}. The design of micro and nano-fluidic devices~\cite{ZN09, zerb07} which mimic biological examples is an emergent field of research with potential applications in medicine and biotechnology~\cite{NatRev}.
If future nano-bot swimmers and pumps might be made through a process of self assembly, the question is how few components are necessary to generate flow, and how simple can the system be.

In this paper we describe a minimal model of pump, inspired by the three-sphere swimmer~\cite{NG04, EP07, GA08, LE+09} where the actuated motion along one axis reduces the description to a one dimension~\cite{CLJ03}. The two-sphere system reduces further its complexity. In the limit of low driving frequency and for average bead separation larger than the bead diameter, the hydrodynamic interaction is described by the Oseen tensor~\cite{MQ99, LE+09}, and the equations of motion are simple enough to allow for explicit calculations. The analytical results back up numerical calculation and experimental data to confirm the surprising result that two beads subject to harmonic potentials can generate a net flow even when the external drive is reciprocal.

\begin{figure}[h!]
\centering
\includegraphics{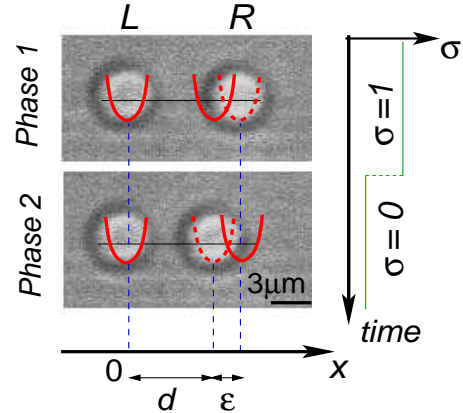}
\caption{(Color online) The two-bead micropump setup. Bead L is held in a stationary optical trap, while bead R is subject to a time-reversible driving force from a switching optical trap. This assembly is capable of generating flow. The harmonic trap potential is shown schematically overlaid on snapshots taken from the experiments.
\label{fig:set-up}}
\end{figure}

The pump is composed of two spheres of radius $a$ labelled with $L$ (Left) and $R$ (Right), positioned on the $x$-axis at an initial distance $d$ as illustrated in FIG.~\ref{fig:set-up}. Each sphere is subject to an harmonic potential, which is realized experimentally by an optical trap, anchored to the laboratory reference frame. Bead $R$ is actively driven: its confining potential switches between two positions along the $x$ direction separated by a distance $\varepsilon$, with a frequency $1/2\tau$. The position of the minimum varies as a function of time  as a a square wave according to $ \sigma(t) = \chi[2n\tau, (2n+1)\tau](t)$, where $\chi$ denotes the indicator function and $n$ is a positive integer. Bead
$L$ is in a stationary potential. There are two distinct phases, corresponding to the values $\sigma = 1, 0$ of the active drive, which constitute a basic cycle of dynamics.

Despite the fact that the trap movement is reciprocal in time, the external actuation by means of springs cannot guarantee an instantaneous balance of the active forces, so the pump is not instantaneously force-free~\cite{Lau07,Lau09}.  Unlike for swimmers, however, this is not problematic: a pump is a spatially confined system and its center of mass lies within a bounded region. In this system this circumstance is also a \emph{necessary condition}, assuring that two degrees of freedom are accessible for pumping. Notably, a weaker condition holds: the pump is force-free on average over a cycle period.

In a low Reynolds number fluid, the actively driven particle undergoes a purely dissipative dynamics, just like a forced oscillator that relaxes towards the minimum of its confining potential. While bead $R$ moves, it interacts hydrodynamically with bead $L$, which can vary its position thanks to the softness of the harmonic potential.  Their dynamics is described, in the regime where $a \ll d$, using the Oseen tensor approximation~\cite{Doi}.  Denoting with $x_L$ the coordinate of $L$ and similarly for $R$, the governing equations read
\begin{equation}
\left\{ \begin{array}{c}
\dot{x}_L = -\omega x_L -\omega \dfrac{\lambda}{r}\Big(x_R -(d-\varepsilon \sigma)\Big);\\
\\
\dot{x}_R = -\omega\Big(x_R -(d-\varepsilon \sigma)\Big) -\omega \dfrac{\lambda}{r} x_L;\\
\end{array}\right.
\label{eq:oseen}
\end{equation}
where we have introduced the three parameters $\gamma=6 \pi \eta a$, $\omega={k}/{\gamma}$, $\lambda={3 a}/{2}$ and the relative distance $r=x_R -x_L $. Here $\eta$ is the fluid viscosity, $\gamma$ is the Stokes' drag coefficient and $k$ is the stiffness of the spring. The equations show that the geometric parameters of the model have a clear interpretation: the inverse of $d$ sets the strength of the hydrodynamic coupling  between the spheres, and $\varepsilon$ is the oscillation amplitude of the actively driven particle.

The model reveals an intriguing property. The presence of two active drives, without any constraint, would provide the system with two ``obvious'' degrees of freedom. However, the temporal dependence of the active mechanism described here is symmetric under time reversal, as can be seen by inspecting FIG.~\ref{fig:set-up} from top to bottom and vice-versa. Thus, at first sight it might appear that the system cannot generate net flow.  Instead, the left-right symmetry is broken.

The pumping can be quantified by focusing on the bead in the resting trap, $L$;  an asymmetry in its trajectory reveals an asymmetry in the flow field. We define the order parameter
\begin{equation}
 \overline{\Delta x}(d, \varepsilon, \tau):= \dfrac{1}{2\tau}\int^{2 \tau}_{0}  x_L dt
\end{equation}
to quantify the magnitude and the direction of the flow as a function of the parameters $d$, $\varepsilon$ and $\tau$.  Its physical interpretation is as follows. Imagine that bead $L$ is an isolated sphere attached to a spring, and subject to the same flow field as the one generated by the pump. Then Hooke's law gives $\overline{F} = k \overline{\Delta x} $, allowing to measure the equivalent mean force exerted by the pump. Using Stokes' law this flow field can be converted into a mean velocity of the fluid $\overline{v} ={\overline{F} }/{\gamma}$.

\begin{figure}[h!]
\centering
\includegraphics{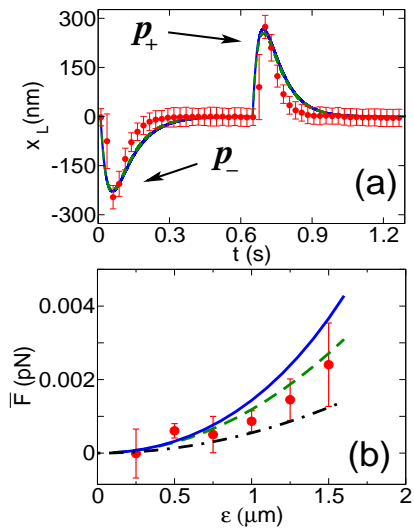}
\caption{(Color online)  Experiment, analytical calculations and simulations agree in quantifying flow generation by the two-bead micropump. (a) As bead R is driven by the optical trap, it causes displacements of bead $L$ around its equilibrium position. Markers show the position of bead L averaged over many cycles. The analytical solution,  (green dashed curve) and simulation data (blue continue curve) closely reproduce the experiments. The displacement values $p_+$ and $p_-$ at
the  peaks are an accurately measurable quantity. Matching the simulations to the peak values enables the observed $p_+$ and $p_-$ to be converted to force by a one-to-one mapping;   (b) The  mean force exerted by the pump on the fluid is measured from the motion of the  bead in the stationary trap, and it grows as a function of the pump stroke length $\varepsilon$.  Markers and lines are as in panel (a). In addition the approximate solution,
Eq.~(\ref{eq:approx-res}) is shown (black dash dotted line).
\label{fig:exp-result}}
\end{figure}

The analysis of the two phases of motion helps to understand how the symmetry breaking occurs~\footnote{For simplicity we consider a relaxed dynamics in which the springs restore completely the spheres to their original position.}. I) in $Phase\,1$, the hydrodynamic coupling between beads $L$ and $R$ has a strength of the order of ${1}/{d}$ and increases as the spheres approach to reach their minimum distance. According to the positions of bead $L$, the fluid is pushed in the sense $x<0$.  Eventually bead $L$ is restored back to equilibrium $x_L=0 $, and in this relaxation some fluid is dragged in the opposite direction. II) in $Phase\,2$, hydrodynamic coupling  is stronger [on the order of ${1}/{(d-\varepsilon)}$]. The dynamics looks similar to the mirror image of the previous one, with bead $R$ dragging bead $L$ in the sense of $x>0$, but now the stronger coupling moves a greater amount of fluid. The overall effect is a net thrust in this direction. An example of such motion for bead $L$ is illustrated in FIG.~\ref{fig:exp-result}(a). The mismatch in hydrodynamic coupling  between end of $Phase\,1$ and beginning of $Phase\,2$ is the root of the symmetry breaking and is made possible by the softness of the driving potentials. Such a phenomenon is analogous to the soft swimming described in~\cite{Lau08}. As we see the pumping direction is determined by the position of the actuated particle: if bead $R$ is active, the pumping is in the direction of $x > 0$ and vice versa if bead $L$ is active.

Introducing the reduced distance $u = r-d$ and mean coordinate $c = (x_L + x_R)/2$, in the approximation of small oscillations the equations of motion can be expanded as power series in the parameter ${u}/{d}$.  With this change, equations~(\ref{eq:oseen}) decouple
into an equation for $u$, independent of $c$, and a linear equation for $c$
involving both variables.
The dynamics at the leading order in $u$ is obtained by substituting
$1/(u+d) \approx 1/d$ in the resulting equations.
This corresponds to the study of the linearized system and a careful
analysis shows that no pumping is achieved.  Physically, this fact can
be understood by interpreting $1/r$
as an effective drag felt by the center of mass $c$. When $r$ is
approximatively constant, then the drag doesn't change during the two
phases of dynamics causing therefore no net thrust on $c$ and thus on
the fluid.

Pumping arises as a non-linear effect which can be seen already at the next order of expansion.
According to $1/(d+u) \approx
1/d-u/d^2$, the reduced distance $u$
has to satisfy of a set of Riccati equations depending on the
parameter $\sigma$:
\begin{equation}
\dot{u} = \mathcal{P}_{\sigma}  + \mathcal{Q}_{\sigma} u + \mathcal{R} u^2,
\end{equation}
where we have defined $\mathcal{P}_{\sigma} := -\varepsilon \sigma \omega\left(1-\lambda/d\right) $, $ \mathcal{Q}_{\sigma} := -\omega \left( 1 -\lambda/d + \sigma (\lambda \varepsilon)/d^2 \right)$ and $\mathcal{R} := -\omega \lambda/d^2 $.
The center of mass equation maintains its linear character,
\begin{equation}
\dot{\tilde{c}}_{\sigma} = -\omega\left(1+\dfrac{\lambda}{d} -\dfrac{u \lambda}{d^2}\right) \tilde{c}_{\sigma}
\end{equation}
for the reduced variable $\tilde{c}_{\sigma} :
=c -\left(d - \varepsilon \sigma \right)/2 $.
Both equations can be solved for $\sigma = 0,1$. Further one must
impose appropriate conditions on the solutions, representing: (i) the
continuity of the whole solution in the middle of the cycle where
$u_0(\tau) = u_1(\tau) $
and $c_0(\tau) = c_1(\tau)$;
 and (ii) the steady state condition for which the positions at the beginning of the cycle coincide with
those at the end, given by
 $u_0(2\tau) = u_1(0)$
and $c_0(2\tau) = c_1(0)$.
 Finally, using
the inverse transformation from $u,c$ to get $x_L$ and $x_R$ and
taking the temporal average of $x_L$, we find that the order parameter
$\overline{\Delta x}$ shows a net pumping over two phases of dynamics.

\begin{figure}[h!]
\centering
\includegraphics{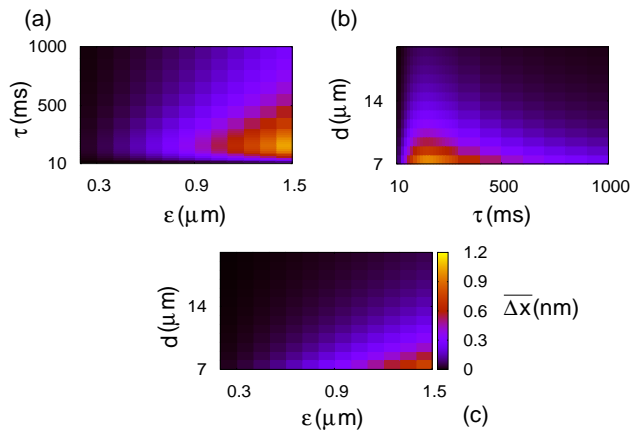}
\caption{(Color online)  Analytical characterization of the pump. (a)
  $\overline{\Delta x}$ as function of $\varepsilon$ and $\tau$. The
  trend as a function of $\varepsilon$ is intuitive: the larger
  the oscillation amplitude, the larger the generated flow. There is a
  non trivial behavior as function of $\tau$, showing that the maximum
  pumping is obtained at maximum $\varepsilon$ for some value of
  $\tau$ in the middle of the scale. $d$ here is fixed to the
  experimental value of $6 \mu$m. (b) $\overline{\Delta x}$ as
  function of $d$ and $\tau$. We see the same phenomenology as in
  (a). Here the effect of the distance is to decrease the pumping
  monotonically. (c) Plot of $\overline{\Delta x}$ as a function of
  $\varepsilon$ and $d$. Again, it shows a monotonic dependence from
  the distance and $\varepsilon$. Looking at these plots,
 is easy to understand that the scaling exponents of
  $\overline{\Delta x}$ are functions of all these parameters. In
  particular, the scaling law of $\overline{\Delta x}$ with
  $\varepsilon$  depends on $\tau$.
In all these cases the values of   $\omega$ and
$a$ are set to typical values taken from the experiments.
\label{fig:surfaces}}
\end{figure}

FIG.~\ref{fig:surfaces} reports the plots of the resulting
expression as functions of $\varepsilon$, $\tau$,
$d$ for typical experimental values of $a$ and $\omega$.  There, the
geometrical variables $\varepsilon$ and $d$ follow the
  straightforward trend where larger oscillation amplitudes
$\varepsilon$ generate larger flow. Also, larger
distance $d$ is
  related to smaller generated flow. However, the plots show also a
non-trivial behavior as a function of the temporal parameter $\tau$.
The form of the solution suggests a natural way to rescale the parameters, defining the nondimensional quantities $\varepsilon^* :=\varepsilon/a$, $d^* := d/a$ and $\tau^*:= \tau \omega$.
There are two different regimes,  i) for small values of $\tau^*$ the
flow increases and ii) for large values of $\tau^*$ it decreases.
The reason for the latter is that in this regime the
beads have sufficient time to relax in the
  harmonic traps, so that the area described by the function
$x_L$ reaches a maximum value, after which it remains constant at
increasing $\tau^*$. Dividing this area by $\tau^*$ gives a decreasing
function of $\tau^*$.
Interestingly,  at the intersection of these two regimes the
plot shows that $\overline{\Delta x}$ presents a maximum, thus
defining an optimal pumping region.  Due to the particular form of the
function $\overline{\Delta x}$, however, this has to be evaluated
numerically.  For the typical experimental values in use it
corresponds to $\tau \approx 165$ms.

For $\tau \gg 1/\omega$, where we can compare with our experimental data, the expression of $\overline{\Delta x}$ simplifies considerably. At the leading order the result is a simple power law dependence
\begin{equation}
\dfrac{\overline{\Delta x}}{a} \approx \dfrac{9}{16 \tau^*}\dfrac{(\varepsilon^*)^2}{(d^*)^3} .
\label{eq:approx-res}
\end{equation}

Integration of the full equations of motion~(\ref{eq:oseen}) is also performed numerically by means of Taylor's method~\cite{CLJ03}. Comparison with the analytical results shows a good agreement, despite the low-order expansion of the analytical solution, as can be seen by looking at FIG.~\ref{fig:exp-result}(a),(b).  This is not surprising as the Oseen tensor is a description valid for of large separation of the beads, and in this regime the perturbative result is a good approximation to the exact solution.

The dynamics has been investigated experimentally by means of an optical tweezer, described fully in~\cite{LE+09}. A laser beam
(IPG Photonics, PYL-1-1064-LP, $\lambda$ = 1064nm, Pmax = 1.1W) is focused
through a water immersion objective (Zeiss, Achroplan IR 63x/0.90 W),
trapping from below. The laser beam is steered via a pair of
acousto-optic deflectors (AA Opto-Electronic, AA.DTS.XY-250@1064nm)
controlled by custom built electronics, allowing multiple trap
generation by time sharing, with sub-nanometer position resolution. Instrument control
and data acquisition (70 frames per second, with an exposure time of 13 ms) are performed by custom software.

A typical experiment consists of trapping two silica beads (3.0$\mu$m diameter, Bangs Labs) in  a solution of glycerol (Fisher, Analysis Grade)
 and is divided in two parts: in the first calibration stage all the traps are kept at rest, and the beads
undergo only Brownian motion confined by the traps.
During the second stage we reproduce the cycle of FIG.~\ref{fig:set-up} with lasers traps, iterating the sequence many times~\cite{LE+09}.  The whole procedure lasts typically 6 minutes, during which we collect about 40~000 frames.  We performed 3 runs for each set of parameters.

By analyzing images by correlation filtering and two dimensional fitting, we obtain the beads' position with subpixel (around 1nm) resolution. The expected values of $\overline{\Delta x}$ are below the limit of the experimental resolution (indeed the smallest forces measured here correspond to 0.1nm displacements), and therefore we characterize the pumping indirectly, relying on the measurable asymmetry of peaks of bead $L$'s position~\cite{LE+09}.
Regarding each cycle as an independent realization of a stochastic process, we construct the mean dynamic cycle of $L$, see FIG.~\ref{fig:exp-result}(a).  Compared to the equilibrium position of the bead, that can be determined from the mean cycle itself but only for large values of $\tau^*$ when the dynamics is fully relaxed, we indicate the
maximum with $p_+$ and the minimum with $p_-$. We define the algebraic sum $\delta := p_+ + p_-$ to quantify the asymmetry of motion and convert it to $\overline{\Delta x}$ with the aid of simulations. This procedure allows to compare extremely small forces, of the order of $5\times 10^{-4}$pN, which, to our knowledge, are the smallest forces measured with optical traps.

In FIG.~\ref{fig:exp-result}(b) we plot the corresponding mean force obtained from the experiments at varying $\varepsilon$ and compare this result with analytical predictions and simulations.
 The experimental values considered are $a =1.5 \mu$m, $d = 6 \mu$m, $\tau=640$ ms, $\omega = 0.022 \pm 0.001 (\mathrm{ms})^{-1}$, $T = 25^o$ C and a trap stiffness value of $k = 5.32 \pm 0.71$ pN/$\mu$m. The results show a good agreement between the measured force and the predicted values. Approximate analytical solution~(\ref{eq:approx-res}) gives a good description of pumping.

 It is interesting to compare the effectiveness of the current minimal pump to the related three-bead model system~\cite{LE+09}.   Close to the optimal pumping region, and for matching values of the stroke $\epsilon$ and inter-particle distance $d$, the average force exerted on the fluid by the three-bead model exceeds the two-bead companion by about one order of magnitude.  The poorer performance with two beads is not surprising, and minimality is obtained at the expense of performance.
However we would like to point out that the two pumps have also a profoundly different nature.  In the three-bead model, pumping is achieved by moving the lateral beads in a non-reciprocal fashion. The direction of the flow is determined  by the first moving trap and can be reverted inverting the trap moves. The nature of the drive in the two-bead model prevents all this and the pumping direction is uniquely determined  by the disposition of the active trap, as discussed above.

In conclusion, an extremely simple system composed of just two spherical beads, only one of which is actuated by a time-reversible trap movement, is shown to be capable of generating flow. A key property of the system is that the beads are held and driven by soft potentials. This allows the two-bead system to explore two degrees of freedom, thus satisfying the ``scallop theorem''. The simplicity of this elementary pump makes it possible to understand  the fluid dynamics analytically, and suggests this as a feasible micro-pump that could be deployed experimentally in the context of microfluidic systems.

\begin{acknowledgments}
One of the authors (M.L.) wishes to acknowledge T.B. Liverpool for many useful discussions.
We acknowledge funding from the Royal Society for an International Joint Project grant.
\end{acknowledgments}


\clearpage
\newpage


\begin{thebibliography}{18}
\expandafter\ifx\csname natexlab\endcsname\relax\def\natexlab#1{#1}\fi
\expandafter\ifx\csname bibnamefont\endcsname\relax
  \def\bibnamefont#1{#1}\fi
\expandafter\ifx\csname bibfnamefont\endcsname\relax
  \def\bibfnamefont#1{#1}\fi
\expandafter\ifx\csname citenamefont\endcsname\relax
  \def\citenamefont#1{#1}\fi
\expandafter\ifx\csname url\endcsname\relax
  \def\url#1{\texttt{#1}}\fi
\expandafter\ifx\csname urlprefix\endcsname\relax\def\urlprefix{URL }\fi
\providecommand{\bibinfo}[2]{#2}
\providecommand{\eprint}[2][]{\url{#2}}

\bibitem[{\citenamefont{Lauga and Powers}(2009)}]{Lau09}
\bibinfo{author}{\bibfnamefont{E.}~\bibnamefont{Lauga}} \bibnamefont{and}
  \bibinfo{author}{\bibfnamefont{T.~R.} \bibnamefont{Powers}},
  \bibinfo{journal}{Reports on Progress in Physics}
  \textbf{\bibinfo{volume}{72}}, \bibinfo{pages}{096601}
  (\bibinfo{year}{2009}).

\bibitem[{\citenamefont{Sherwood}(2001)}]{Sher}
\bibinfo{author}{\bibfnamefont{L.}~\bibnamefont{Sherwood}},
  \emph{\bibinfo{title}{Human Physiology: From Cells to Systems}}
  (\bibinfo{publisher}{Brooks/Cole}, \bibinfo{year}{2001}).

\bibitem[{\citenamefont{Bray}(2001)}]{Bray}
\bibinfo{author}{\bibfnamefont{D.}~\bibnamefont{Bray}},
  \emph{\bibinfo{title}{Cell movements: From Molecules to Motility}}
  (\bibinfo{publisher}{Garland Publishing}, \bibinfo{year}{2001}).

\bibitem[{\citenamefont{Purcell}(1977)}]{Pur77}
\bibinfo{author}{\bibfnamefont{E.}~\bibnamefont{Purcell}},
  \bibinfo{journal}{American Journal of Physics} \textbf{\bibinfo{volume}{45}},
  \bibinfo{pages}{3} (\bibinfo{year}{1977}).

\bibitem[{\citenamefont{Raz and Avron}(2007)}]{RA07}
\bibinfo{author}{\bibfnamefont{O.}~\bibnamefont{Raz}} \bibnamefont{and}
  \bibinfo{author}{\bibfnamefont{J.~E.} \bibnamefont{Avron}},
  \bibinfo{journal}{New Journal of Physics} \textbf{\bibinfo{volume}{9}},
  \bibinfo{pages}{437} (\bibinfo{year}{2007}).

\bibitem[{\citenamefont{Najafi and Golestanian}(2004)}]{NG04}
\bibinfo{author}{\bibfnamefont{A.}~\bibnamefont{Najafi}} \bibnamefont{and}
  \bibinfo{author}{\bibfnamefont{R.}~\bibnamefont{Golestanian}},
  \bibinfo{journal}{Phys Rev E Stat Nonlin Soft Matter Phys}
  \textbf{\bibinfo{volume}{69}}, \bibinfo{pages}{062901}
  (\bibinfo{year}{2004}).

\bibitem[{\citenamefont{Leoni et~al.}(2009)\citenamefont{Leoni, Kotar,
  Bassetti, Cicuta, and Lagomarsino}}]{LE+09}
\bibinfo{author}{\bibfnamefont{M.}~\bibnamefont{Leoni}},
  \bibinfo{author}{\bibfnamefont{J.}~\bibnamefont{Kotar}},
  \bibinfo{author}{\bibfnamefont{B.}~\bibnamefont{Bassetti}},
  \bibinfo{author}{\bibfnamefont{P.}~\bibnamefont{Cicuta}}, \bibnamefont{and}
  \bibinfo{author}{\bibfnamefont{M.~C.} \bibnamefont{Lagomarsino}},
  \bibinfo{journal}{Soft Matter} \textbf{\bibinfo{volume}{5}},
  \bibinfo{pages}{472} (\bibinfo{year}{2009}).

\bibitem[{\citenamefont{Zargar et~al.}(2009)\citenamefont{Zargar, Najafi, and
  Miri}}]{ZN09}
\bibinfo{author}{\bibfnamefont{R.}~\bibnamefont{Zargar}},
  \bibinfo{author}{\bibfnamefont{A.}~\bibnamefont{Najafi}}, \bibnamefont{and}
  \bibinfo{author}{\bibfnamefont{M. F.}~\bibnamefont{Miri}},
  \bibinfo{journal}{Physical Review E (Statistical, Nonlinear, and Soft Matter
  Physics)} \textbf{\bibinfo{volume}{80}}, \bibinfo{pages}{026308}
  (\bibinfo{year}{2009}).

\bibitem[{\citenamefont{Leigh and Zerbetto}(2007)}]{zerb07}
\bibinfo{author}{\bibfnamefont{D.}~\bibnamefont{Leigh}} \bibnamefont{and}
  \bibinfo{author}{\bibfnamefont{F.}~\bibnamefont{Zerbetto}},
  \bibinfo{journal}{Angewandte Chemie International Edition}
  \textbf{\bibinfo{volume}{46}}, \bibinfo{pages}{72} (\bibinfo{year}{2007}).

\bibitem[{\citenamefont{Dittrich and Manz}(2006)}]{NatRev}
\bibinfo{author}{\bibfnamefont{P.~S.} \bibnamefont{Dittrich}} \bibnamefont{and}
  \bibinfo{author}{\bibfnamefont{A.}~\bibnamefont{Manz}}, \bibinfo{journal}{Nat
  Rev Drug Discov} \textbf{\bibinfo{volume}{5}}, \bibinfo{pages}{210}
  (\bibinfo{year}{2006}).

\bibitem[{\citenamefont{Earl et~al.}(2007)\citenamefont{Earl, Pooley, Ryder,
  Bredberg, and Yeomans}}]{EP07}
\bibinfo{author}{\bibfnamefont{D.~J.} \bibnamefont{Earl}},
  \bibinfo{author}{\bibfnamefont{C.~M.} \bibnamefont{Pooley}},
  \bibinfo{author}{\bibfnamefont{J.~F.} \bibnamefont{Ryder}},
  \bibinfo{author}{\bibfnamefont{I.}~\bibnamefont{Bredberg}}, \bibnamefont{and}
  \bibinfo{author}{\bibfnamefont{J.~M.} \bibnamefont{Yeomans}},
  \bibinfo{journal}{The Journal of Chemical Physics}
  \textbf{\bibinfo{volume}{126}}, \bibinfo{pages}{064703}
  (\bibinfo{year}{2007}).

\bibitem[{\citenamefont{Golestanian and Ajdari}(2008)}]{GA08}
\bibinfo{author}{\bibfnamefont{R.}~\bibnamefont{Golestanian}} \bibnamefont{and}
  \bibinfo{author}{\bibfnamefont{A.}~\bibnamefont{Ajdari}},
  \bibinfo{journal}{Physical Review E (Statistical, Nonlinear, and Soft Matter
  Physics)} \textbf{\bibinfo{volume}{77}}, \bibinfo{pages}{036308}
  (\bibinfo{year}{2008}).

\bibitem[{\citenamefont{Lagomarsino et~al.}(2003)\citenamefont{Lagomarsino,
  Jona, and Bassetti}}]{CLJ03}
\bibinfo{author}{\bibfnamefont{M.~C.} \bibnamefont{Lagomarsino}},
  \bibinfo{author}{\bibfnamefont{P.}~\bibnamefont{Jona}}, \bibnamefont{and}
  \bibinfo{author}{\bibfnamefont{B.}~\bibnamefont{Bassetti}},
  \bibinfo{journal}{Phys. Rev. E} \textbf{\bibinfo{volume}{68}},
  \bibinfo{pages}{021908} (\bibinfo{year}{2003}).

\bibitem[{\citenamefont{Meiners and Quake}(1999)}]{MQ99}
\bibinfo{author}{\bibfnamefont{J.-C.} \bibnamefont{Meiners}} \bibnamefont{and}
  \bibinfo{author}{\bibfnamefont{S.~R.} \bibnamefont{Quake}},
  \bibinfo{journal}{Phys. Rev. Lett.} \textbf{\bibinfo{volume}{82}},
  \bibinfo{pages}{2211} (\bibinfo{year}{1999}).

\bibitem[{\citenamefont{Lauga}(2007)}]{Lau07}
\bibinfo{author}{\bibfnamefont{E.}~\bibnamefont{Lauga}},
  \bibinfo{journal}{Physical Review E (Statistical, Nonlinear, and Soft Matter
  Physics)} \textbf{\bibinfo{volume}{75}}, \bibinfo{pages}{041916}
  (\bibinfo{year}{2007}).

\bibitem[{\citenamefont{Doi and Edwards}(1986)}]{Doi}
\bibinfo{author}{\bibfnamefont{M.}~\bibnamefont{Doi}} \bibnamefont{and}
  \bibinfo{author}{\bibfnamefont{S.}~\bibnamefont{Edwards}},
  \emph{\bibinfo{title}{The Theory of Polymer Dynamics}}
  (\bibinfo{publisher}{Oxford University Press}, \bibinfo{year}{1986}).

\bibitem[{\citenamefont{Trouilloud et~al.}(2008)\citenamefont{Trouilloud, Yu,
  Hosoi, and Lauga}}]{Lau08}
\bibinfo{author}{\bibfnamefont{R.}~\bibnamefont{Trouilloud}},
  \bibinfo{author}{\bibfnamefont{T.~S.} \bibnamefont{Yu}},
  \bibinfo{author}{\bibfnamefont{A.~E.} \bibnamefont{Hosoi}}, \bibnamefont{and}
  \bibinfo{author}{\bibfnamefont{E.}~\bibnamefont{Lauga}},
  \bibinfo{journal}{Physical Review Letters} \textbf{\bibinfo{volume}{101}},
  \bibinfo{pages}{048102} (\bibinfo{year}{2008}).

\bibitem[{\citenamefont{Ashkin}(1997)}]{Ash97}
\bibinfo{author}{\bibfnamefont{A.}~\bibnamefont{Ashkin}},
  \bibinfo{journal}{Proc Natl Acad Sci U S A} \textbf{\bibinfo{volume}{94}},
  \bibinfo{pages}{4853} (\bibinfo{year}{1997}).


\end{thebibliography}
\end{document}